\documentclass[]{aa}  

\usepackage{natbib}
\bibpunct{(}{)}{;}{a}{}{,}
\usepackage{graphicx}
\usepackage{revsymb}	
\usepackage{amsmath}	
\usepackage[usenames]{color}	
\usepackage{epstopdf}	
\usepackage{txfonts}
\usepackage{amssymb}
\usepackage{color}
\usepackage[justification=centering]{caption}
\usepackage{geometry} 
\usepackage[parfill]{parskip} 
\usepackage{amssymb}
\usepackage{slantsc}
\usepackage{array}
\usepackage{url}
\usepackage{amsfonts}
\usepackage[colorlinks=true,linkcolor=blue, urlcolor=blue, citecolor=blue]{hyperref}
\usepackage{sidecap}
\usepackage{graphicx}
\usepackage{xcolor}
\usepackage{nicefrac}
\usepackage{subfigure}
\usepackage{epsfig}
\colorlet{rouge}{red!70!darkgray}

\begin{document}
\title{Constraints on the properties of macroscopic transport in the Sun from combined lithium and beryllium depletion}
\author{G. Buldgen\inst{1} \and A. Noels\inst{1} \and A. M. Amarsi\inst{2} \and D. Nandal\inst{3} \and C. Pezzotti\inst{1}\and R. Scuflaire\inst{1} \and M. Deal \inst{4} \and N. Grevesse \inst{1,5}}
\institute{STAR Institute, Université de Liège, Liège, Belgium \and Theoretical Astrophysics, Department of Physics and Astronomy, Uppsala University, Box 516, 751 20 Uppsala, Sweden \and University of Virginia, Astronomy Building, 530 McCormick Road, P.O. Box 400325, Charlottesville, VA 22904 \and LUPM, Universit\'e de Montpellier, CNRS, Place Eug\`ene Bataillon, 34095 Montpellier, France \and Centre Spatial de Liège, Université de Liège, Angleur-Liège, Belgium}
\date{July, 2024}
\abstract{The Sun is a privileged laboratory of stellar evolution, thanks to the quality and complementary nature of available constraints. Using these observations, we are able to draw a detailed picture of its internal structure and dynamics which form the basis of the successes of solar modelling. Amongst such constraints, the depletion of lithium and beryllium are key tracers of the required efficiency and extent of macroscopic mixing just below the solar convective envelope. Thanks to revised determinations of these abundances, we may use them in conjunction with other existing spectroscopic and helioseismic constraints to study in detail the properties of macroscopic transport.}
{We aim at constraining the efficiency of macroscopic transport at the base of the convective envelope and determining the compatibility of the observations with a suggested candidate linked with the transport of angular momentum in the solar radiative interior.}
{We use recent spectroscopic observations of lithium and beryllium abundance and include them in solar evolutionary model calibrations. We test the agreement of such models in terms of position of the convective envelope, helium mass fraction in convective zone, sound speed profile inversions and neutrino fluxes.}
{We constrain the required efficiency and extent of the macroscopic mixing at the base of the solar convective envelope, finding that a power law of density with an index $n$ between 3 and 6 would reproduce the data, with efficiencies at the base of the envelope of about $6000\; \rm{cm^{2}/s}$, depending on the value of n. We also confirm that macroscopic mixing worsens the agreement with neutrino fluxes and that the current implementations of the magnetic Tayler instability are unable to explain the observations.}{}
\titlerunning{Macroscopic transport from lithium and beryllium depletion} 
\maketitle
\section{Introduction}

The Sun is a crucial laboratory of stellar physics at microscopic and macroscopic scales and an important reference point for stellar evolution in general \citep[see][and references therein]{JCD2021}. Thanks to helioseismic and spectroscopic constraints, supplemented by the detections of neutrinos directly informing us on the properties of the solar core, we are able to draw a detailed picture of the internal structure and dynamics of the Sun \citep[see e.g.][and ref therein]{JCD91Conv,BasuYSun, JCD1996, Basu97BCZ,SchouRota,BasuSun,OrebiGann2021,Appel2022,Basilico2023}. Apart from a few studies \citep{Buldgen2023,Buldgen2024,Basinger2024}, most of the recent standard solar models have yet to take into account the effects of light element depletion \citep[e.g.][]{Vinyoles2017}. Indeed, the current photospheric lithium and beryllium abundances inform us on the dynamical processes acting at the base of the solar convective envelope and had already been recognized as crucial constraints in earlier studies of the Sun \citep[e.g.][]{Pinsonneault1989,Swenson1991,JCD1996,Richard1996,Charbonneau2000,Brun02,Boesgaard2009,JCD2018,Jorgensen2018,Boesgaard2022} and other stars \citep[e.g.][for reviews]{Boesgaard1988,Lyubimkov2016,Salaris2017,Lyubimkov2018}. Lithium and beryllium are key tracers of macroscopic mixing at the base of the convective envelope of cool stars thanks to their low fusion temperatures ($2.5\times 10^{6} \rm{K}$ and $3.5\times 10^{6} \rm{K}$, respectively). Both elements have been studied in detail and linked with the effects of rotation (or under a more general umbrella term, macroscopic mixing) \citep{Pinsonneault1997, Chaboyer1998}. The link between lithium depletion and age has also been clearly demonstrated in numerous cases \citep[e.g.][for recent works]{Carlos2016,Baraffe2017,Thevenin2017,Carlos2019,Dumont2021}. The question of lithium has also been discussed in the context of metal poor stars and primordial nucleosynthesis models, where the amount of mixing in old metal-poor stars  plays a central role \citep[see e.g.][and refs therein]{Deal2021,Deal2021b} and globular clusters are prime targets to constrain primordial nucleosynthesis \citep[see e.g.][for recent works]{Boesgaard2023,Boesgaard2024}.

The solar photospheric values have changed significantly over the past thirty years. These abundances are provided using the usual logarithmic scale relative to hyddrogen, defined as $\rm{A(X)}=\log (N(X)/N(H))+12.00$, with $\rm{A(H)=12}$.  For lithium, \citet{Anders1989} report a value of $\rm{A(Li)}=1.16$, \citet{Grevesse1998} report $1.10$ while \citet{Asplund2009} report $\rm{A(Li)}=1.05\pm0.10$. Even more recently, \citet{Wang2021} estimate $\rm{A(Li)}=0.96\pm 0.05$, which was adopted in \citet{Asplund2021}.  In general, the steady decrease in lithium is largely due to improved modelling of departures from local thermodynamic equilibrium (LTE). For beryllium, the picture is more complex, with the evolution going from $\rm{A(Be)}=1.15$ in \citet{Anders1989} based on \citet{Chmielewski1975}, to $\rm{A(Be)}=1.40$ in \citet{Grevesse1998} and $\rm{A(Be)}=1.38$ in \citet{Asplund2009}.  The latter two results are based on the analyses of \citet{Balachandran1998} and \citet{Asplund2004} that carefully calibrated the missing UV opacity but did not take into account a significant blend of the \ion{Be}{II} resonance line.  Taking this into account, combined with 3D non-LTE models, \citet{Amarsi2024} find $\rm{A(Be)}=1.21 \pm 0.05$.

In this work, we take advantage of recent present-day photospheric determinations of \citet{Wang2021} and \citet{Amarsi2024} combined with expectations of the present-day surface abundances from meteorites, converted to the solar scale and taking into account the effects of the transport of chemicals, $\rm{A(Li)} = 3.27\pm0.03$ and $\rm{A(Be)} = 1.31\pm0.04$ \citep{Lodders2021} to study in detail the properties of macroscopic transport of chemicals at the base of the solar convective zone (BCZ). The latter serve as initial abundances for our models, to which is added the effect of mixing and solar calibration (see Sect. \ref{Sec:Models}). We also study in detail its potential link with angular momentum transport based on implementations of the combined effects of the shear instability, the meridional circulation and the magnetic Tayler instability \citep{Eggenberger2022} as applied in the solar case to explain the observed lithium depletion. The recently determined beryllium abundance by \citet{Amarsi2024} demonstrates the need for mixing also on the main sequence, as the higher fusion temperature of beryllium does not allow it to be efficiently burned during the pre main-sequence. It essentially offers a new anchoring point to study the required properties of macroscopic mixing at the base of the solar convective zone in the context of the current issues following the revision of solar abundances.

We start by introducing our solar evolutionary models in Sect. \ref{Sec:Models}, presenting a calibration of macroscopic mixing at the base of the convective envelope and discussing its implication for classical helioseismic inversions and neutrino fluxes. We detail the observed behaviour of our models regarding global properties in Sect. \ref{Sec:Models}. We study various approaches to represent macroscopic mixing at the base of the convective envelope and discuss our findings in light of potential candidates mentioned in previous works \citep{Eggenberger2022}. We further discuss these aspects in Sect. \ref{Sec:MagneticInst} when looking at the expected behaviour of the magnetic Tayler instability (called Tayler-Spruit instability of Tayler-Spruit dynamo). We conclude in Sect. \ref{Sec:Conc} with further progress avenues and application of our findings in constructing data-driven models of the Sun as well as extensions to other stars to place our findings in the more general context of the modelling of solar-like stars and the determination of their age using asteroseismic data. 

\section{Solar models}\label{Sec:Models}

We compute solar calibrated models using the Liège Stellar Evolution Code \citep[CLES,][]{ScuflaireCles}, including the latest version of the SAHA-S equation of state \citep{Gryaznov04,Baturin}, the OPAL opacities \citep{OPAL} supplemented at low temperatures by tables of \citet{Ferguson}, and the \citet{Asplund2021} abundances (hereafter AAG21). The models were calibrated following a standard calibration procedure using the initial hydrogen mass fraction, X, the initial heavy element mass fraction, Z, and the mixing length parameter, $\alpha_{\rm{MLT}}$ as free parameters and the solar radius, luminosity and surface metallicy $(\rm{Z/X})_{s}$ as constraints. The surface metallicity for the calibration was taken from \citet{Asplund2021}, $\rm{Z/X})_{s}=0.0187$ and the solar radius and luminosity taken from the 2015 IAU recommended values \citep{Mamajek2015}. The solar age was fixed to $4.57$ Gy in all our calibrations. Microscopic diffusion was included in the models using the formalism of \citet{Thoul} and the screening coefficients of \citet{Paquette}. Those effects include both thermal diffusion, gravitional settling and composition diffusion. Radiative accelerations are not included in these models as they have been shown to have a very limited impact in solar interior conditions \citep{Turcotte1998}. The pre main-sequence (PMS) phase is fully taken into account in our evolutionary computations, starting from a fully convective seed of one solar mass, following its initial contraction and the onset of reactions such as Deuterium burning and out-of-equilibrium $^{3}\rm{He}$ burning. We do not, however, consider accretion of protosolar material as in \citet{Kunitomo2022} but this has been shown to have limited impact on the lithium and beryllium depletion.

The evolution during the pre-main-sequence leads to a significant depletion of lithium and a small depletion of beryllium. Not considering it would bias our conclusions regarding the efficiency of the turbulent transport during the main-sequence. For example, our models show a depletion of about $0.8$ dex for lithium during the PMS phase. An even higher depletion is found if overshooting at the base of the convective zone is included, while the impact on beryllium remains limited, as a result of its higher fusion temperature. It is likely that aiming at reproducing both lithium, beryllium and the helioseismic position of the BCZ using overshooting alone is unfeasible and leaves only turbulent mixing to deplete beryllium during the main-sequence. Further evidence of this can be seen in Fig. 6 of \citet{Eggenberger2022} and in Fig. 1 of \citet{Buldgen2023}, where the lithium abundance of young solar twins is not reproduced when overshooting is used to replace the base of the convective zone at the helioseismic value. 

Additional macroscopic mixing was modelled using two approaches, first we used the \citet{Proffitt1991} empirical coefficient, that is a simple power law of density
\begin{align}
D^{n}_{D_{T}}=D_{T} \left(\frac{\rho_{\rm{BCZ}}}{\rho}\right)^{n} \label{eq:Proff}
\end{align}
with $\rho_{\rm{BCZ}}$ the density at the BCZ position, $\rho$ the local density, $D_{T}$ and $n$ being the free parameters in this formalism. \\

The second parametrization of macroscopic mixing is based on an asymptotic formulation of the transport by the combined effect of shear instability, meridional circulation and the magnetic Tayler instability. Recently, \citet{Eggenberger2022TS} proposed a simple modification of the parametrization of the magnetic Tayler instability, from which we can generalize the diffusion coefficient found in \citet{Buldgen2024},

\begin{align}
D^{n}_{C_{T}}=\rm{D_{h}}\frac{1}{C^{2}_{T}}\left( \Omega \right)^{3n/2}\left( N_{\mu}  \right)^{-n/2}\left( \frac{\eta_{M}}{r^{2}} \right)^{-n/2} \label{eq:Magn}
\end{align}
with $\eta_{M}$ the magnetic diffusivity, $\Omega$ the angular velocity in the upper solar radiative zone (fixed to the helioseismic value), $N_{\mu}$ the mean molecular weight term of the Brunt-Väisälä frequency, $\rm{D_{h}}$ the horizontal turbulence and $C_{T}$ a calibration coefficient to take into account the uncertainties on the damping timescale of the azimuthal magnetic field.  

These coefficients are then adjusted to reproduce the observed lithium \citep{Wang2021} and beryllium \citep{Amarsi2024} depletions but are not included directly in the calibration scheme. In the case of the formulation linked with the magnetic instability, the only adjustable parameter is the efficiency of the mixing, linked to $\rm{D_{h}}$, which is adjusted on lithium as in \citet{Eggenberger2022}. 

Given that the efficiency of this mixing is directly linked to the position of the base of the convective zone, we also investigate the required variations in efficiency for two ways of replacing the position of the BCZ at the helioseismic value, following \citep{Buldgen2024}. First, we use adiabatic overshooting in an extended calibration procedure. Second, we use an ad-hoc increase of opacity in radiative regions just under the BCZ. The two methods lead to different depletion histories of lithium and different signatures in helioseismic inversions (See Sect.\ref{Sec:HelioNeut}). 

\subsection{Light element depletion and helium mass fraction}\label{Sec:Parameters}

We start by taking a look at the relevant global properties describing solar models, namely the radial coordinate position of the base of the convective envelope, the mass coordinate at the position of the base of the convective envelope, the helium mass fraction in the convective envelope, the photospheric lithium and beryllium abundances as well as the initial helium abundance and mixing length parameter $\alpha_{\rm{MLT}}$ used for the calibration. The values of these various properties for each model are provided in Table \ref{tabModelsDTurb}. We also recall the observational constraints available from the literature for some of these quantities. The names of the models reflect the required mixing parameters in Eq. \ref{eq:Proff}. The initial abundances for lithium and beryllium are taken from the meteoritic abundances from \citet{Lodders2021}, corrected by the effects of mixing which lead to a slightly higher initial abundances by about $0.04$ dex depending on the exact solar calibration procedure.

It is evident that by calibrating the efficiency of the macroscopic transport at the BCZ on the lithium depletion, we always reach very similar values of $\rm{Y_{CZ}}$ and $\left(r/R\right)_{\rm{BCZ}}$. This confirms our previous findings regarding the issue of models including macroscopic mixing and helioseismic constraints. Namely that the BCZ position is pushed out by the erasing of the metallicity peak observed in standard solar models \citep[see][for a discussion of the origin of this peak]{Baturin2006}.

\begin{table*}[h]
\caption{Global parameters of the solar evolutionary models including turbulent diffusion as treated in Eq. \ref{eq:Proff}}
\label{tabModelsDTurb}
  \centering
\begin{tabular}{r | c | c | c | c | c | c | c | c }
\hline \hline
\textbf{Name}&\textbf{$\left(r/R\right)_{\rm{BCZ}}$}&\textbf{$\left( m/M \right)_{\rm{CZ}}$}&\textbf{$\mathit{Y}_{\rm{CZ}}$} & A(Li) & A(Be) & $\mathit{Y}_{0}$& $\alpha_{\rm{MLT}}$ & $\mathit{Z}_{0}$\\ \hline
Model D$^{2}_{3500}$&$0.7257$&$0.9792$& $0.2479$ & $1.17$ & $1.05$ & $0.264$ & $1.979$ & $0.0145$\\
Model D$^{2}_{3800}$&$0.7258$&$0.9793$& $0.2481$ & $1.08$ & $1.02$ & $0.264$ & $1.978$ & $0.0145$\\ 
Model D$^{2}_{4000}$&$0.7259$&$0.9793$& $0.2481$ & $1.02$ & $1.01$ & $0.264$ & $1.978$ & $0.0145$\\ 
Model D$^{2}_{4500}$&$0.7260$&$0.9793$& $0.2484$ & $0.87$ & $0.98$ & $0.264$ & $1.977$ & $0.0145$\\
Model D$^{2}_{5500}$&$0.7261$&$0.9793$& $0.2487$ & $0.60$ & $0.91$ & $0.264$ & $1.975$ & $0.0145$\\
Model D$^{3}_{6500}$&$0.7256$&$0.9792$& $0.2477$ & $1.06$ & $1.14$ & $0.264$ & $1.980$ & $0.0145$\\ 
Model D$^{3}_{6700}$&$0.7257$&$0.9792$& $0.2477$ & $1.02$ & $1.13$ & $0.264$ & $1.980$ & $0.0145$\\
Model D$^{3}_{7500}$&$0.7257$&$0.9792$& $0.2479$ & $0.90$ & $1.11$ & $0.264$ & $1.980$ & $0.0145$\\
\hline
Observations &$0.713\pm0.001^{1}$& $/$& $0.2485\pm0.0035^{2}$ & $0.96\pm0.05^{3}$& $1.31\pm0.04^{4}$& $/$ & $/$ &$/$\\
\hline
\end{tabular}

\small{\textit{Note:} $^{1}$ \citet{Basu97BCZ}, $^{2}$ \citet{BasuYSun}, $^{3}$\citet{Wang2021},$^{4}$ \citet{Amarsi2024}}
\end{table*}

\begin{figure*}
	\centering
		\includegraphics[width=17cm]{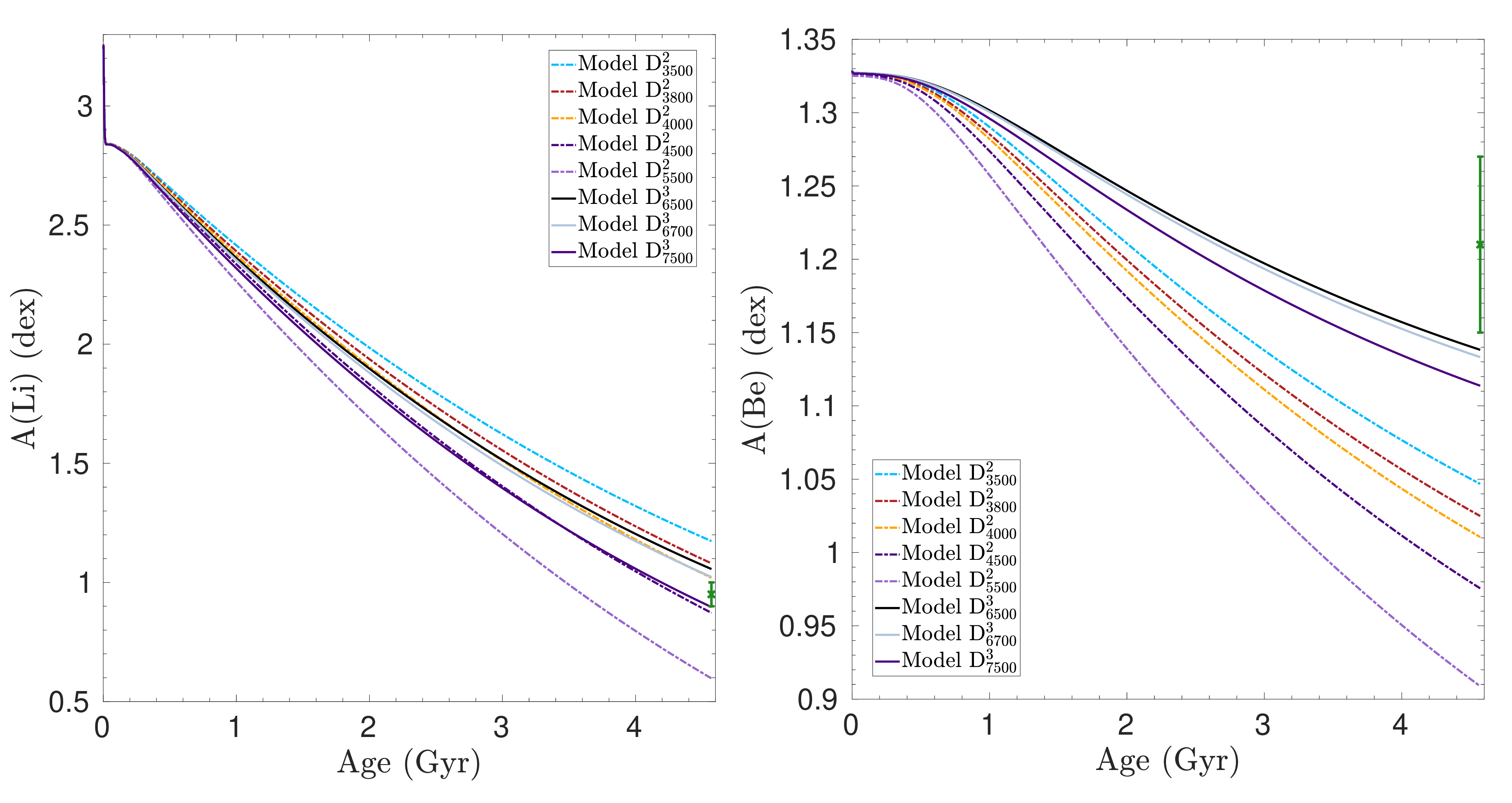}
	\caption{Left panel: lithium depletion as a function of age for the solar models in Table \ref{tabModelsDTurb}. Right panel: beryllium depletion as a function of age for the solar models in Table \ref{tabModelsDTurb}. The green crosses indicate the observed values of lithium \citep{Wang2021} and beryllium \citep{Amarsi2024}. The duration of the PMS phase in our models is of $40$ Myr and leads to a significant depletion of lithium and a moderate depletion of beryllium that are unfortunately hardy visible.}
		\label{Fig:LiBeDT}
\end{figure*}

In Fig. \ref{Fig:LiBeDT}, we illustrate the beryllium and lithium depletion for the various solar-calibrated models of Table \ref{tabModelsDTurb}. The models were computed to show a range of predicted photospheric abundances around the \citet{Wang2021} lithium  value. From a quick inspection, we can clearly see that a value of $n\geqslant 3$ will be required to allow to reproduce both lithium and beryllium simultaneously. Indeed, all models with $n=2$, represented by dashed lines in Fig. \ref{Fig:LiBeDT} have a too high depletion of beryllium, even if we reduce the multiplicative factor $D_{T}$ to a value leading to a too low depletion of lithium. The situation is less problematic for models with $n=3$, as the models come very close to agreeing to both lithium and beryllium simultaneously. This already raises some issues regarding the initial calibrations of \citet{Eggenberger2022} which used $n=1.3$ to best reproduce the effects of angular momentum transport processes.

\begin{figure}
	\centering
		\includegraphics[width=9cm]{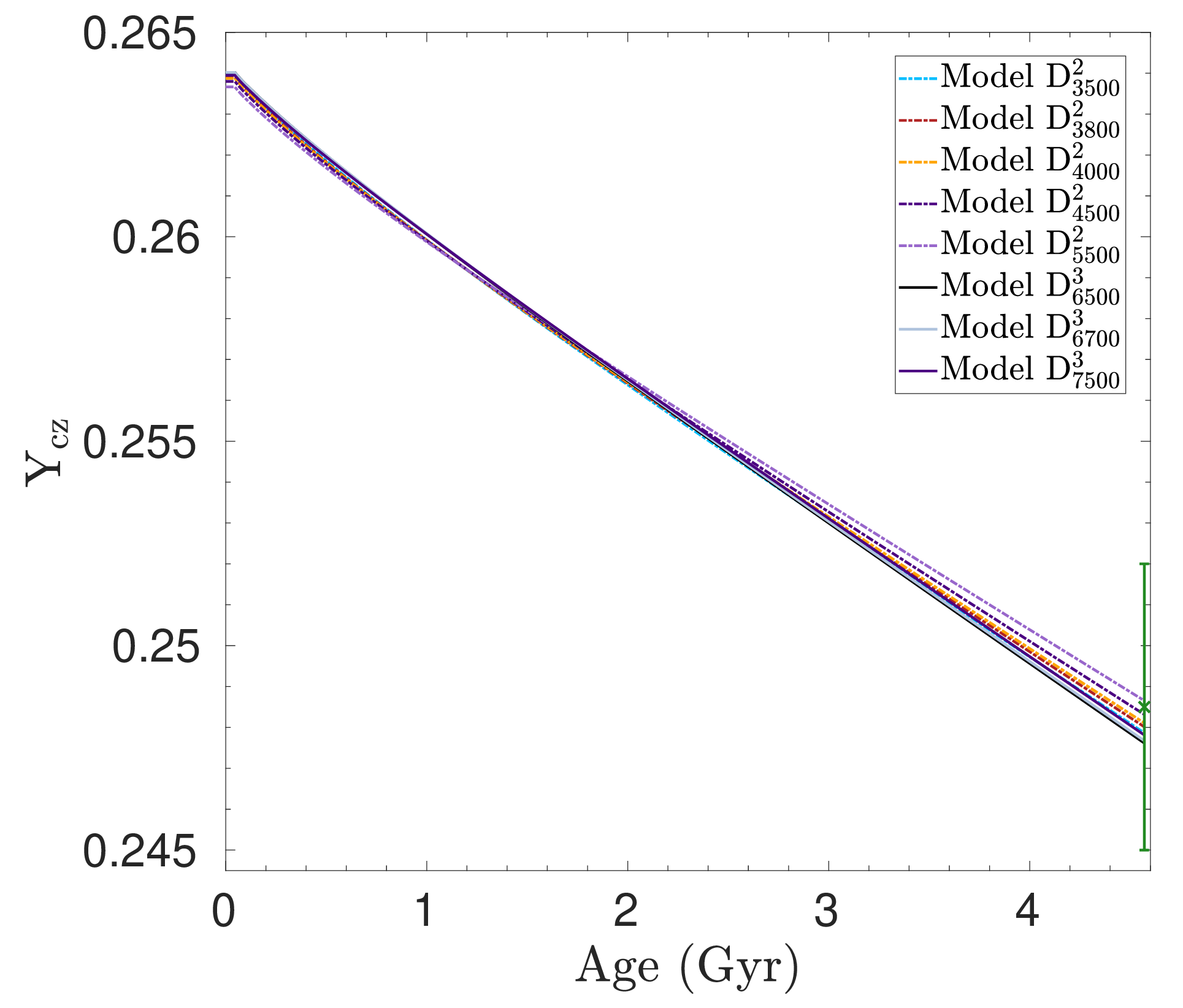}
	\caption{Evolution of the surface helium abundance as a function of age for the models of Table \ref{tabModelsDTurb}. The green crosses indicate the surface helium abundance inferred from helioseismology by \citet{BasuYSun}}
		\label{Fig:YDT}
\end{figure}

In Fig. \ref{Fig:YDT}, we illustrate the evolution of the surface helium abundance as a function of age for the models of Table \ref{tabModelsDTurb}. We can see that the required efficiency to reproduce both lithium and beryllium leads to a reduction of the initial helium abundance as well as a lower efficiency of settling over time. This reduction is significant as it allows models built with the AAG21 abundances to reproduce the surface helium inferred from helioseismology by \citet{BasuYSun}. This reduced efficiency can be understood from the shape of the diffusion velocity curves close to the BCZ \citep[see][]{Turcotte} and agrees with previous findings \citep{Richard1996,Dumont2021,Buldgen2023}. 

In Table \ref{tabModelsMagnetic}, we provide the same constraints for models implementing an asymptotic formulation of the combined effect of shear instability, meridional circulation and magnetic Tayler instability. The classical formulation of \citet{spr02} had been put forward in \citet{Eggenberger2022} as a potential explanation for the lithium depletion as a result of the shear in the solar radiative envelope being reduced to an appropriate value. This formulation was however shown to lead to a too high depletion of beryllium at the age of the Sun compared to the observations \citep{Amarsi2024}. This is further confirmed from the supplementary materials of \citet{Eggenberger2022} (see their Fig. S3), which show that their mixing coefficient of chemicals goes as low as $0.4$R$_{\odot}$, exhibiting a relatively flat behaviour.

However, as mentioned in \citet{Buldgen2024} and noted in multiple works \citep{Cantiello2014,Deheuvels2020}, another key issue of the classical formalism of the Tayler-Spruit dynamo is its inability to reproduce the internal rotation of solar-like subgiants and red giants. Revision of the formalism by \citet{Fuller2019} and an ad-hoc recalibration of the efficiency of the coefficient for the dissipation of the magnetic fields ($C_{T}$) in \citet{Eggenberger2022TS} have been suggested to increase the efficiency of the angular momentum transport after the main sequence. In what follows, the impact of such revisions on the solar properties is studied in detail. We start by providing the global parameters of the solar models in Table \ref{tabModelsMagnetic}. The exponents in Eq. \ref{eq:Magn} are varied following \citet{Eggenberger2022TS}, $n=3$ being linked with the \citet{Fuller2019}, where \citet{Eggenberger2022TS} assumed two modifications of $C_{T}$ to marginally improve the agreement with post-main sequence rotation, namely $C_{T}=1$ or $0.125$ and $n=1$ and $C_{T}=256$ being their suggested recalibration. From Table \ref{tabModelsMagnetic}, we can already grasp the main problem of models including a recalibrated version of the magnetic Tayler instability, namely that their behaviour is that of a standard solar model. Indeed, the main impact of these modified formalisms is to lock the allowed shear below a given threshhold, provided by Eq. 12 in \citet{Eggenberger2022TS}. Therefore when using the recalibrated formalisms with the same horizontal turbulence efficiency, the allowed threshhold is so low that no mixing of chemicals occur during the main-sequence.  

\begin{table*}[h]
\caption{Global parameters of the solar evolutionary models including macroscopic mixing as treated in Eq. \ref{eq:Magn}}
\label{tabModelsMagnetic}
  \centering
\begin{tabular}{r | c | c | c | c | c | c | c | c }
\hline \hline
\textbf{Name}&\textbf{$\left(r/R\right)_{\rm{BCZ}}$}&\textbf{$\left( m/M \right)_{\rm{CZ}}$}&\textbf{$\mathit{Y}_{\rm{CZ}}$} & A(Li) & A(Be) & $\mathit{Y}_{0}$& $\alpha_{\rm{MLT}}$ & $\mathit{Z}_{0}$\\ \hline
Model D$^{1}_{1}$       &$0.7263$&$0.9794$& $0.2491$ & $0.97$ & $0.77$ & $0.264$ & $1.972$ & $0.0144$\\
Model D$^{1}_{256}$     &$0.7223$&$0.9785$& $0.2398$ & $2.74$ & $1.33$ & $0.268$ & $2.021$ & $0.0151$\\ 
Model D$^{3}_{1}$       &$0.7223$&$0.9785$& $0.2398$ & $2.74$ & $1.33$ & $0.268$ & $2.021$ & $0.0151$\\
Model D$^{3}_{0.125}$   &$0.7220$&$0.9784$& $0.2398$ & $2.74$ & $1.33$ & $0.268$ & $2.021$ & $0.0151$\\
Model OP D$^{1}_{1}$    &$0.7138$&$0.9762$& $0.2516$ & $0.97$ & $0.86$ & $0.266$ & $2.000$ & $0.0144$\\ 
Model OP D$^{1}_{256}$  &$0.7127$&$0.9758$& $0.2427$ & $2.64$ & $1.33$ & $0.267$ & $2.039$ & $0.0150$\\
Model OP D$^{3}_{1}$    &$0.7126$&$0.9760$& $0.2427$ & $2.64$ & $1.33$ & $0.267$ & $2.039$ & $0.0145$\\ 
Model OP D$^{3}_{0.125}$&$0.7127$&$0.9758$& $0.2427$ & $2.63$ & $1.33$ & $0.267$ & $2.039$ & $0.0150$\\
Model Ov D$^{1}_{1}$    &$0.7133$&$0.9768$& $0.2477$ & $0.96$ & $1.05$ & $0.264$ & $1.983$ & $0.0150$\\
Model Ov D$^{1}_{256}$  &$0.7133$&$0.9768$& $0.2401$ & $2.23$ & $1.33$ & $0.267$ & $2.018$ & $0.0150$\\
\hline
Observations &$0.713\pm0.001^{1}$& $/$& $0.2485\pm0.0035^{2}$ & $0.96\pm0.05^{3}$& $1.31\pm0.04^{4}$& $/$ & $/$ & $/$\\
\hline
\end{tabular}

\small{\textit{Note:} $^{1}$ \citet{Basu97BCZ}, $^{2}$ \citet{BasuYSun}, $^{3}$\citet{Wang2021},$^{4}$ \citet{Amarsi2024}}
\end{table*}

\begin{figure*}
	\centering
		\includegraphics[width=17cm]{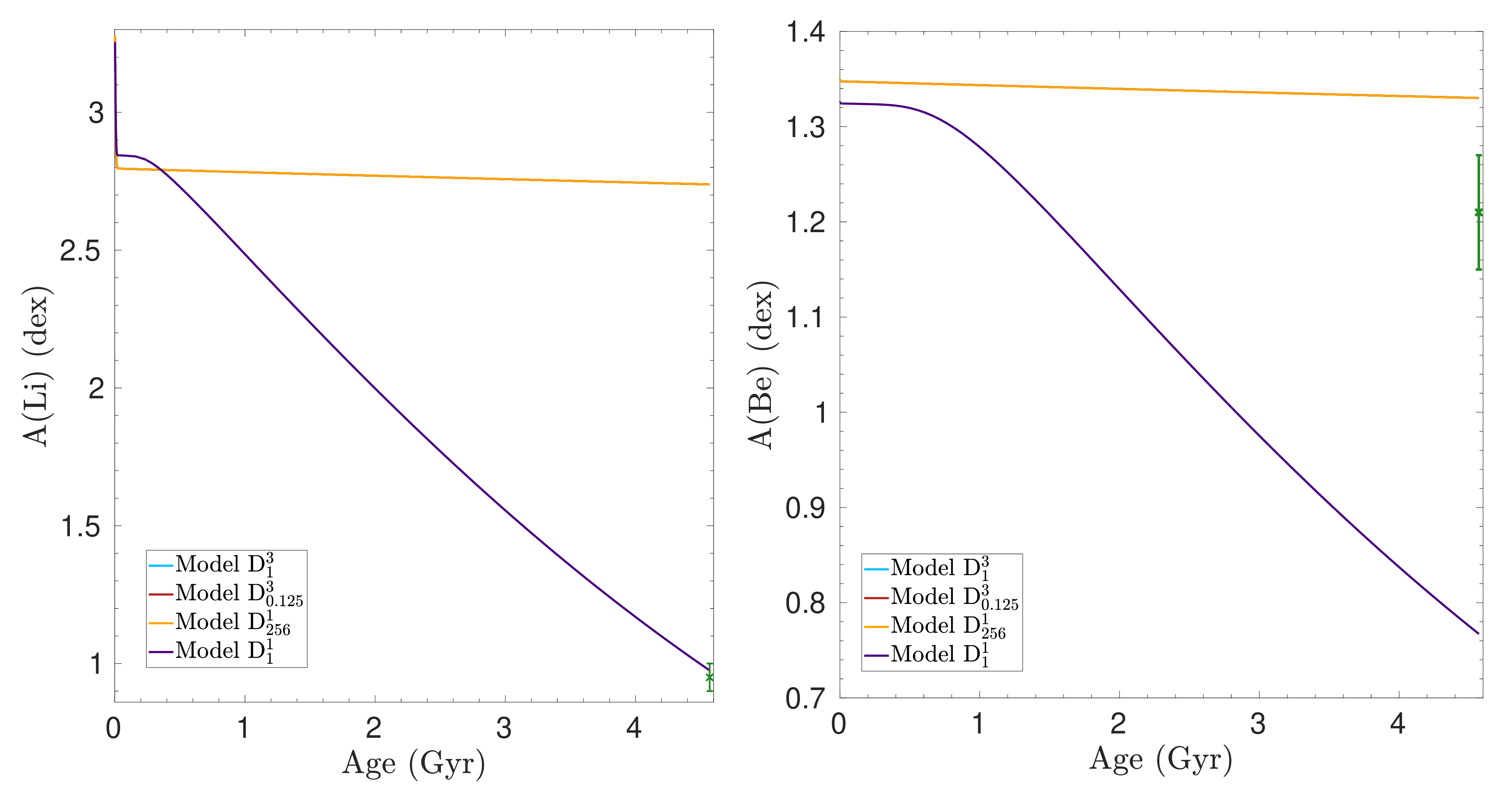}
	\caption{Left panel: lithium depletion as a function of age for models including an asymptotic treatment of the magnetic instabilities. Right panel: beryllium depletion as a function of age for models including an asymptotic treatment of the magnetic instabilities. In both panels, models using the revised treatment of magnetic instabilities, namely Model D$^{3}_{0.125}$, Model D$^{3}_{1}$ and Model D$^{1}_{256}$ are indistinguishable due to their identical behaviours.}
		\label{Fig:LiBeD9}
\end{figure*}

The evolution of the photospheric lithium and beryllium abundances is illustrated in Fig. \ref{Fig:LiBeD9} for the various models including the asymptotic treatment of macroscopic mixing by shear, circulation and magnetic instabilities. A clear pattern emerges again, from this analysis, namely that the need for a very efficient angular momentum transport to reproduce the core rotation of red giants pushes the models with this implementation of macroscopic transport to behave like standard models and show no depletion of light elements.

\begin{figure*}
	\centering
		\includegraphics[width=17cm]{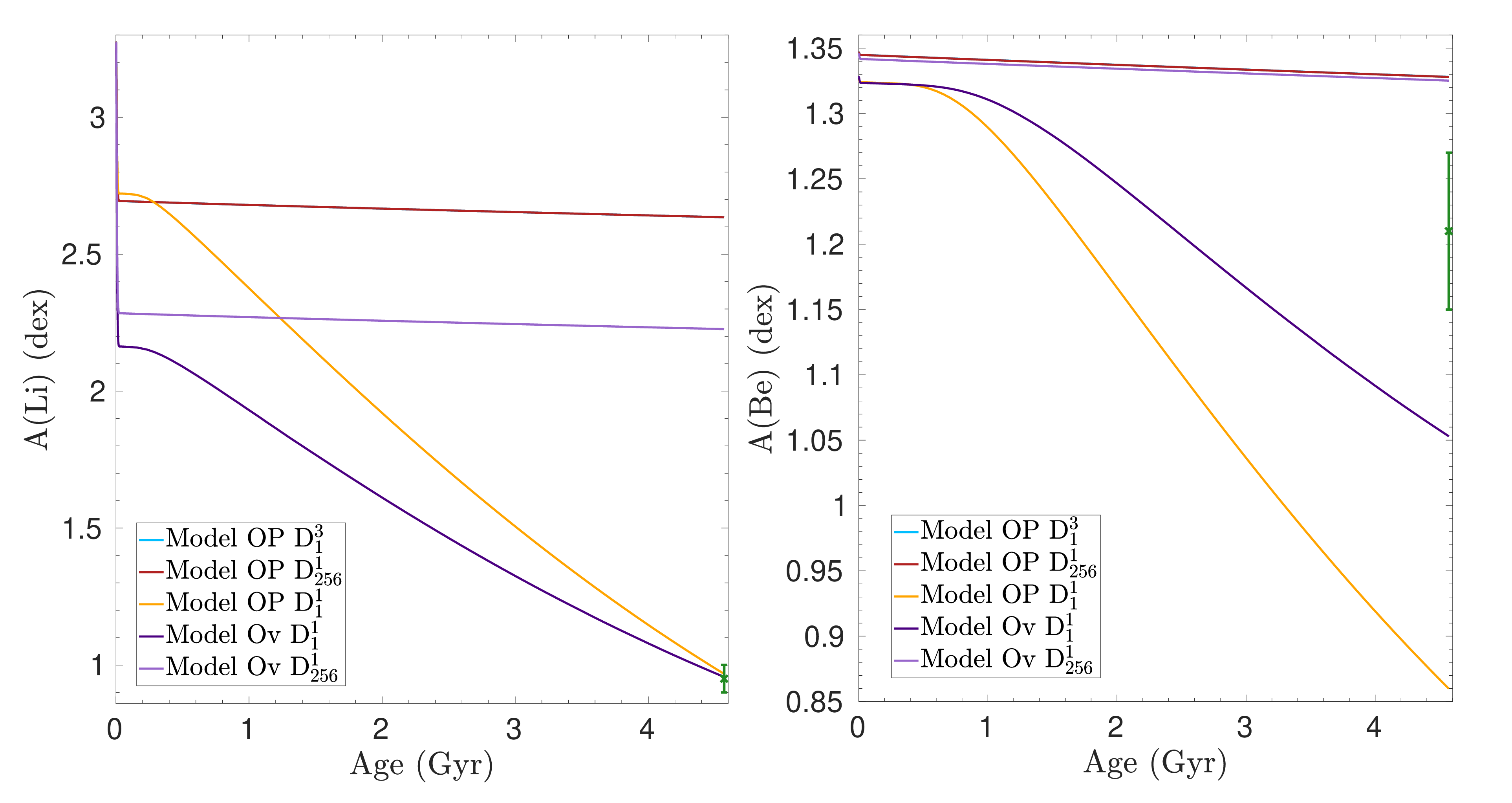}
	\caption{Left panel: lithium depletion as a function of age for models including an asymptotic treatment of the magnetic instabilities coupled with overshooting or opacity increase. Right panel: beryllium depletion as a function of age for models including an asymptotic treatment of the magnetic instabilities coupled with overshooting or opacity increase. In both panels, models using the revised treatment of magnetic instabilities, namely Model OP D$^{1}_{256}$, Model OP D$^{3}_{1}$ and Model OP D$^{3}_{0.125}$ are undistinguishable due to their identical behaviours.}
		\label{Fig:LiBeD9Op}
\end{figure*}

Further attempts were made including overshooting and an opacity increase replacing the BCZ at the helioseismic value to attempt to find a regime where such models could reproduce simultaneously the lithium and beryllium depletion. The results for these attempts are illustrated in Fig. \ref{Fig:LiBeD9Op}, again, even if we reduce the efficiency of the mixing and increase the lithium depletion during the PMS as a result of envelope overshooting, the mixing induced by the combined effects of magnetic instabilities, shear and circulation goes too deep and leads to a too high depletion of beryllium at the age of the Sun. This confirms the trends observed in \citet{Amarsi2024} on a larger set of models. We will come back to the physical implications of these results in Sect. \ref{Sec:MagneticInst}. 

As mentioned above, the position of the BCZ matters significantly for the calibration of macroscopic transport. Indeed, the overall mixing efficiency will likely be overestimated in AAG21 models that show a significant mismatch in their position of the BCZ. A thinner envelope leads to lower temperatures at the BCZ and thus requires a more efficient  mixing to deplete lithium and beryllium. In previous works \citep[e.g.][]{Buldgen2024}, we discussed the impact of using overshooting or an opacity increase to replace the BCZ position in models including macroscopic mixing. As shown in \citet{Buldgen2023}, including adiabatic overshooting to correct the BCZ in solar models leads to a too high depletion of lithium in young solar twins, therefore we chose here to focus on the effect of an opacity increase in our models. We summarize in Table \ref{tabModelsOPAC} the main properties of these models. As can be seen, our evolutionary models are tailored to reproduce the BCZ position, and are in good agreement with the helium abundance inferred from helioseismology. The required opacity increase was modelled in the same way as in \citet{Buldgen2024}, with an amplitude of about 14$\%$ close to the BCZ. As we can see from the naming convention of the models, the efficiency of the mixing for a given $n$ is approximately half of what was required in Table \ref{tabModelsDTurb}, this is a direct result of the change in the position of the BCZ as a function of time. 

\begin{table*}[h]
\caption{Global parameters of the solar evolutionary models reproducing both Lithium, Beryllium and BCZ position.}
\label{tabModelsOPAC}
  \centering
\begin{tabular}{r | c | c | c | c | c | c | c | c }
\hline \hline
\textbf{Name}&\textbf{$\left(r/R\right)_{\rm{BCZ}}$}&\textbf{$\left( m/M \right)_{\rm{CZ}}$}&\textbf{$\mathit{Y}_{\rm{CZ}}$} & A(Li) & A(Be) & $\mathit{Y}_{0}$& $\alpha_{\rm{MLT}}$ & $\mathit{Z}_{0}$\\ \hline
Model OP D$^{3}_{3400}$ &$0.7136$&$0.9761$& $0.2497$ & $1.09$ & $1.17$ & $0.267$ &$2.009$ & $0.0145$\\
Model OP D$^{3}_{3700}$ &$0.7136$&$0.9761$& $0.2498$ & $1.00$ & $1.16$ & $0.267$ &$2.009$ & $0.0145$\\
Model OP D$^{4}_{6000}$ &$0.7135$&$0.9761$& $0.2496$ & $0.93$ & $1.21$ & $0.267$ &$2.010$ & $0.0145$\\
Model OP D$^{4}_{6400}$ &$0.7135$&$0.9761$& $0.2499$ & $0.93$ & $1.20$ & $0.267$ &$2.009$ & $0.0145$\\
Model OP D$^{5}_{8800}$ &$0.7134$&$0.9760$& $0.2494$ & $0.98$ & $1.25$ & $0.267$ &$2.011$ & $0.0145$\\
Model OP D$^{5}_{9000}$ &$0.7134$&$0.9760$& $0.2494$ & $0.96$ & $1.25$ & $0.267$ &$2.011$ & $0.0145$\\ 
Model OP D$^{6}_{14000}$&$0.7133$&$0.9760$& $0.2492$ & $0.97$ & $1.27$ & $0.267$ &$2.012$ & $0.0145$\\ 
\hline
Observations &$0.713\pm0.001^{1}$& $/$& $0.2485\pm0.0035^{2}$ & $0.96\pm0.05^{3}$& $1.31\pm0.04^{4}$& $/$ & $/$ & $/$\\
\hline
\end{tabular}

\small{\textit{Note:} $^{1}$ \citet{Basu97BCZ}, $^{2}$ \citet{BasuYSun}, $^{3}$\citet{Wang2021},$^{4}$ \citet{Amarsi2024}}
\end{table*}

\begin{figure*}
	\centering
		\includegraphics[width=17cm]{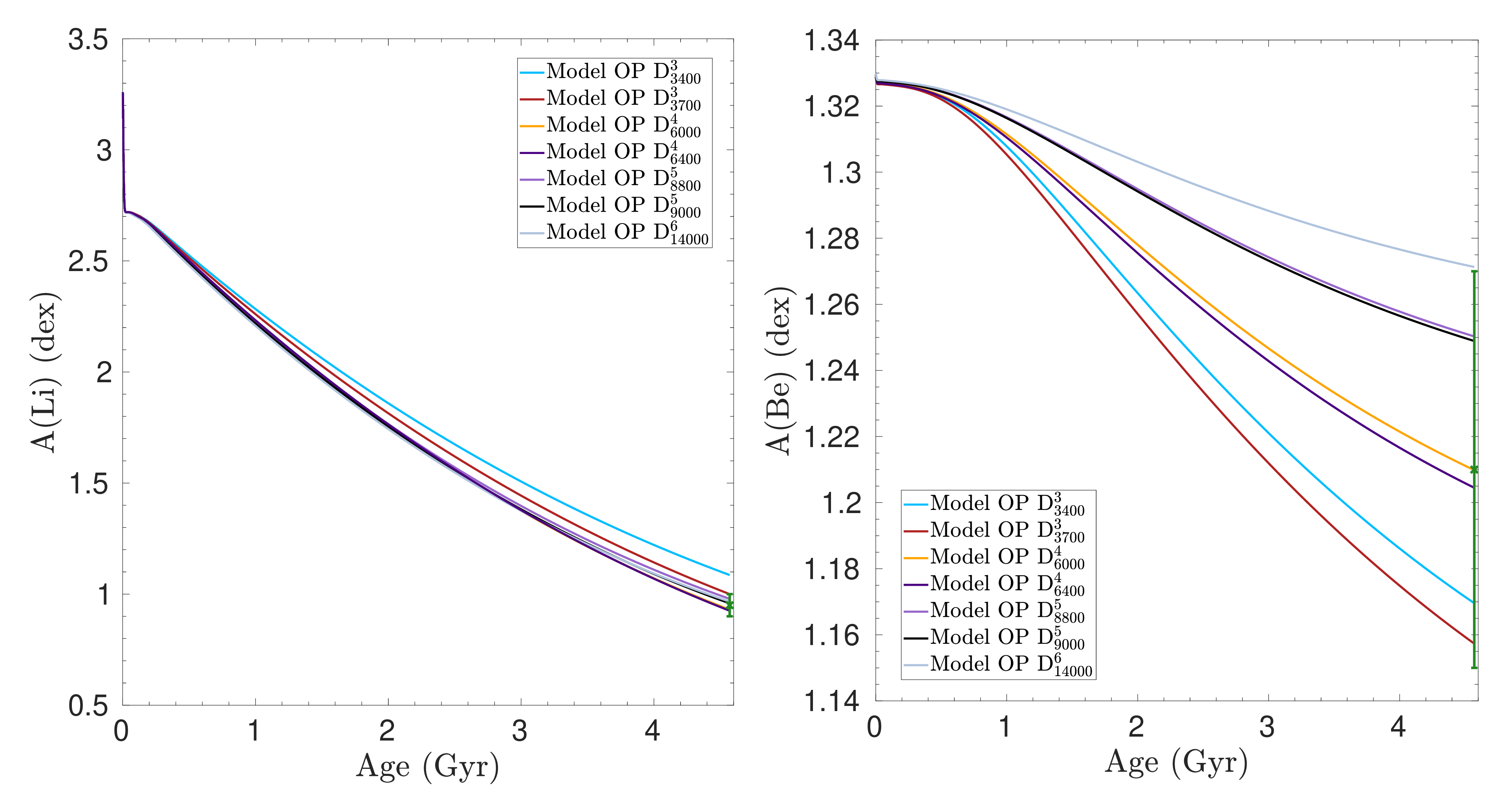}
	\caption{Left panel: lithium depletion as a function of time for the models of Table \ref{tabModelsOPAC} including an opacity increase at the BCZ. Right panel: beryllium depletion as a function of time for the models of Table \ref{tabModelsOPAC}. The green crosses indicate the observed values at the current solar age.}
		\label{Fig:LiBeDTOp}
\end{figure*}

In Fig. \ref{Fig:LiBeDTOp}, we illustrate the lithium and beryllium depletion as a function of time for the models of Table \ref{tabModelsOPAC}. By allowing to play on both parameters of Eq. \ref{eq:Proff}, reproducing both constraints becomes quite straightfoward, and we see that agreeing with the beryllium abundance at the solar age also leads to a much narrower evolution of lithium as a function of time. From the observed trends in the behaviour of the mixing coefficient, it appears that allowed values of $n$ range between 3 and 6, with the lower values leading to a slightly too high depletion of beryllium and the highest ones to almost no depletion at all. This is in line with the calibrated values found in A type stars with n=3 or 4 \citep{Richer2000,Richard2001,Michaud2005}. Further extending such analyses to solar-like stars observed with the \textit{Kepler} mission for which high-quality asteroseismic data is available might provide more insight on the underlying physical process responsible for lithium depletion as well as further test its link with angular momentum transport mechanisms.  

\subsection{Constraints on macroscopic transport efficiency}\label{Sec:Efficiency}

\begin{figure*}
	\centering
		\includegraphics[width=17cm]{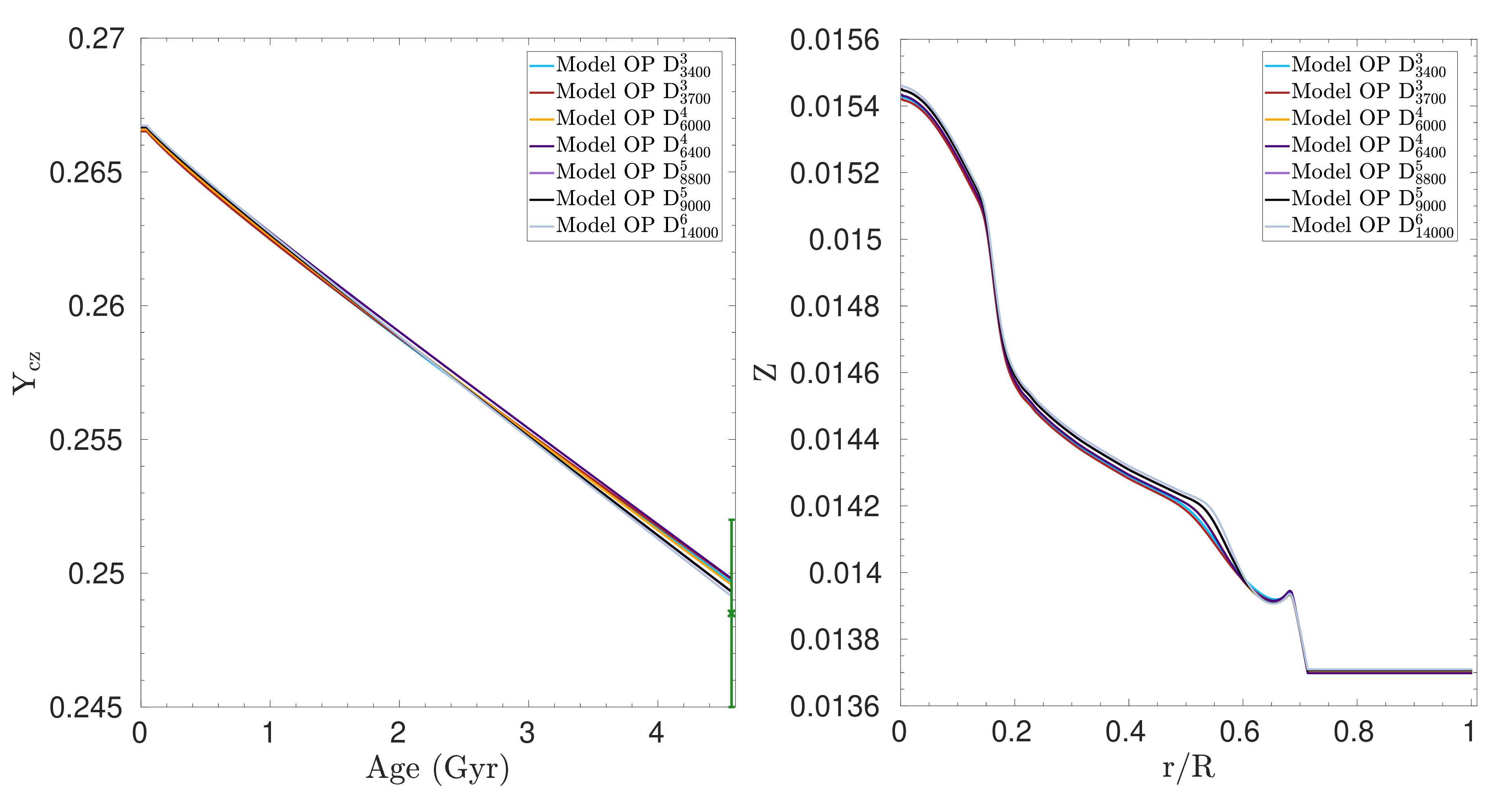}
	\caption{Left panel: surface helium mass fraction evolution as a function of time for our models including macroscopic transport and opacity modifications. Right panel: metallicity profile as a function of normalized radius for the our models including macroscopic transport and opacity modifications.}
		\label{Fig:YZOpDT}
\end{figure*}

In Fig. \ref{Fig:YZOpDT}, we illustrate the evolution of surface helium mass fraction as a function of time and the metallicity profile of our models reproducing both lithium and beryllium photospheric abundances and the BCZ position through an opacity increase. The first striking feature of these models is the very narrow range of initial and final helium abundance they exhibit. A small trend is observed with sharper coefficients (higher values of n) leading to slightly lower final helium values as a result of the shallower mixing regions that allows for a larger impact of settling on the evolution. This is a direct result of the calibration of the mixing using beryllium that essentially controls the efficiency required during the main sequence. This implies that a relatively narrow range of final and initial helium abundances are allowed when macroscopic mixing is included, allowing to refine previous estimates of protosolar helium based on standard solar models \citep{Serenelli2010}.

A second important feature of the models are the similarities in the metallicity profile close to the BCZ. Essentially, the presence of efficient transport in these layers leads to the metals being efficiently mixed down to $0.6$R$_{\odot}$. Some form of differentiation between the various efficiencies appears below these layers, where the effects of settling may still be felt. Nevertheless this spread remains small and the same conclusion can be drawn looking at the central metallicity values. These results have strong implications on neutrino fluxes, as we will see in Sect. \ref{Sec:HelioNeut} and are a natural result of models including macroscopic mixing.

\subsection{Impact on helioseismic results and neutrino fluxes}\label{Sec:HelioNeut}
\begin{figure}
	\centering
		\includegraphics[width=9cm]{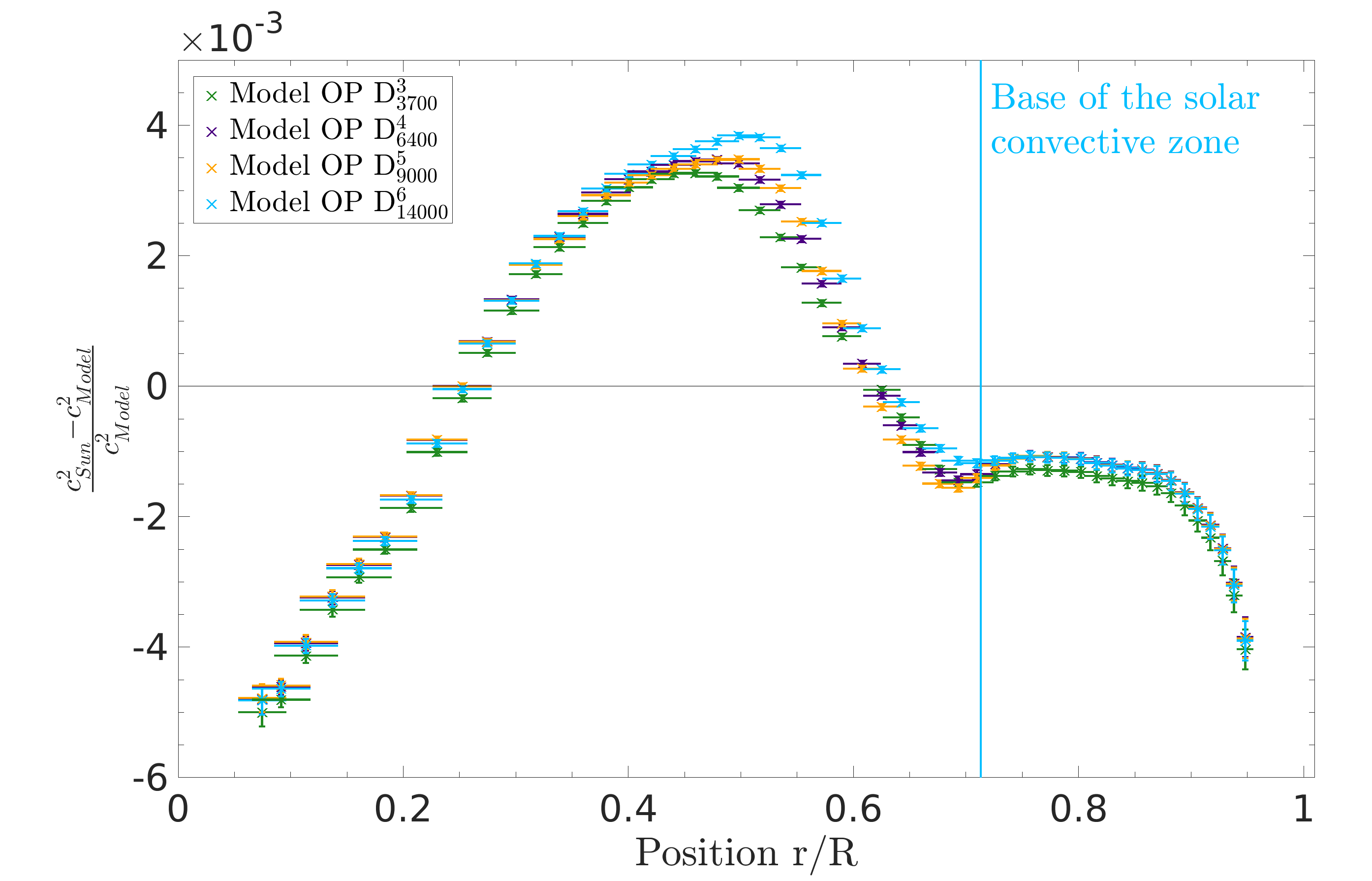}
	\caption{Relative squared adiabatic sound speed differences between the Sun and models using the OPAL and OPLIB opacities, either within the standard solar model framework or including macroscopic mixing of chemical elements.}
		\label{Fig:SoundSpeedOpac}
\end{figure}

We illustrate in Fig. \ref{Fig:SoundSpeedOpac} the sound speed profile for our models including macroscopic mixing at the BCZ and an opacity increase in its vicinity to replace its position at the helioseismic value. As can be seen, the usual deviation at the BCZ has been completely erased by the combination of the opacity increase and the macroscopic mixing. Some deviations nevertheless remain in the radiative layers and in the core. Such effects have been seen in previous studies \citep[see e.g.][]{Buldgen2023,Buldgen2024} and can be attributed to the lowering of the core metallicity that results from the effects of macroscopic mixing on the calibration. As mentioned above, the effects are similar to helium and thus a lower initial metallicity is found for models including macroscopic mixing compared to standard solar models only considering the effects of settling. It is no surprise that the localized increase in opacity we implemented, following \citet{Buldgen2024}, does not correct the lower layers. We stress again that the opacity modification implemented here is not based on a physical process, we merely implemented a Gaussian-like increase of about $14 \%$ in the vicinity of the iron peak around $\log \rm{T}=6.35$, similarly to previous extended calibrations experiments \citep{Ayukov2017,Kunitomo2021}. It is likely that some of the effects being currently studied \citep{Pain2020,Pradhan2024,Pradhan2024b} would also impact higher temperature regions in the solar models. Therefore Fig. \ref{Fig:SoundSpeedOpac} mostly illustrates how the BCZ region can be easily corrected and how model including macroscopic transport show very similar properties in their deeper radiative layers, which was already hinted at from looking at Fig. \ref{Fig:YZOpDT}.  

The situation is similar for neutrino fluxes regarding the comparative behaviours of the models that are provided in Table \ref{tabNeutrinos}. Given their similar chemical evolution and initial chemical composition resulting from the solar calibration, they have quite similar central conditions, and therefore neutrino fluxes. However, in this case, macroscopic mixing leads to much lower fluxes for $\phi(\rm{Be})$, $\phi(\rm{B})$ and $\phi(\rm{CNO})$, therefore to a stark disagreement with observed values if we focus on the Borexino results. The situation is not so problematic if one considers the values provided by the work of \citet{OrebiGann2021} based on multiple experiments. This is also in agreement with what we found in \citet{Buldgen2024} for models using the abundances of \citet{Magg2022}. This issue is more pronounced for the AAG21 models as they start from a lower initial value due to their intrinsically lower metallicity. It remains to be seen whether other processes such as planetary formation \citep{Kunitomo2022} might compensate the lowering of the central metallicity or whether other effects may enter into play. Indeed, planetary formation, through the evolution of the metallicity of the accreted matter from the protosolar disk may significantly impact the central metallicity at the current solar age and lead to an increase of about $5\%$. Further tests with latest nuclear reaction rates \citep{Acharya2024} and effects of dynamical electronic screening may need to be investigated. It should also be mentioned that potential opacity revisions \citep{Bailey,Pain2020,Pradhan2024,Pradhan2024b} may also extend to higher temperature and affect the initial conditions of solar calibrations and thus the predicted neutrino fluxes of solar models. In addition, a steeper macroscopic transport coefficient seems to also slightly reduce the impact on neutrino fluxes, although by a very small amount. 
\begin{table*}[h]
\caption{Neutrino fluxes of the evolutionary models}
\label{tabNeutrinos}
  \centering
\begin{tabular}{r | c | c | c | c }
\hline \hline
\textbf{Name}& $\phi(\rm{pp})$ $\left[ \times 10^{10}/\rm{cm}^{2}/\rm{s}\right]$& $\phi(\rm{Be})$ $\left[ \times 10^{9}/\rm{cm}^{2}/\rm{s}\right]$&$\phi(\rm{B})$ $\left[ \times 10^{6}/\rm{cm}^{2}/\rm{s}\right]$&$\phi(\rm{CNO})$ $\left[ \times 10^{8}/\rm{cm}^{2}/\rm{s}\right]$\\ \hline
Model OP D$^{3}_{3700}$& $6.00$ & $4.61$ & $4.73$ &$3.82$\\
Model OP D$^{4}_{6400}$& $5.99$ & $4.61$ & $4.72$ & $3.83$\\ 
Model OP D$^{5}_{8800}$& $6.00$ & $4.61$ & $4.74$ & $3.84$\\
Model OP D$^{6}_{14600}$& $6.00$ & $4.62$ & $4.75$ & $3.84$\\
O-G21$^{1}$ & $5.97^{+0.0037}_{-0.0033}$ & $4.80^{+0.24}_{-0.22}$ & $5.16^{+0.13}_{-0.09}$ & $-$\\
Borexino$^{2}$ & $6.1^{+0.6}_{-0.7}$ & $4.99^{+0.13}_{-0.14}$ & $5.68^{+0.39}_{-0.41}$ & $6.7^{+1.2}_{-0.8}$\\
\hline
\end{tabular}

\small{\textit{Note:} $^{1}$ \citet{OrebiGann2021}, $^{2}$ \citet{Borexino2018}, \citet{Borexino2020}, \citet{Basilico2023}}
\end{table*}

\section{Link with magnetic Tayler instability}\label{Sec:MagneticInst}

As discussed above, the magnetic Tayler instability is a potential candidate, alongside fossil magnetic fields \citep{Gough1998} and internal gravity waves \citep{Charbonnel2005, Pincon2016}, to explain the current solar rotation profile. While some debate existed in the literature regarding its potential occurrence in hydrodynamical simulations \citep{Zahn2007}, recent works by \citet{Petitdemange2023} showed that it was actually present in their simulations. Revision of the work of \citet{spr02} was carried out by \citet{Fuller2019} to explain the slow rotation of red giants and core-helium burning stars. 

From Figs. \ref{Fig:LiBeD9} and \ref{Fig:LiBeD9Op}, we can see that there seems to be an intrinsic problem with the macroscopic transport provided by the asymptotic depiction of rotation and the magnetic Tayler instability. In the ``classical'' case of the initial \citet{spr02} formulation, the obtained beryllium depletion is too high. This is a direct result of the functional expression of the transport coefficient that goes as deep as $0.4\rm{R}_{\odot}$ and only has a slow decreasing trend. Indeed, \citet{Eggenberger2022} show that its efficiency could be reproduced in CLES models using $n=1.3$ in the \citet{Proffitt1991} formulation. The reason for this efficiency is linked with the allowed shear by the magnetic Tayler instability. Even if the rotation profile appears to be quite flat, the instability is considered to be active only after a given level of shear is reached \citep[see][for a discussion]{spr02}, meaning that it undergoes some sort of limit cycle that still allows for some mixing of chemicals at the base of the convective zone, where mean molecular weight gradients are formed over time as a result of microscopic diffusion and the envelope slows down as a result of the magnetic braking of the surface.

The main issue with this formulation of the magnetic Tayler instabilities is that it leads to a too low efficiency of the angular momentum transport on the subgiant and red giant branch and is therefore unable to reproduce the asteroseismic constraints for subgiants \citep{Deheuvels2014, Deheuvels2020} and red giants \citep{Cantiello2014}. This led to the derivation of a new formulation by \citet{Fuller2019} and an ad-hoc calibration of the efficiency of the magnetic field dissipation in \citet{Eggenberger2022TS}. Both formulations provide a similar agreement on the red giant branch but fail to reproduce the trend observed in core rotations at the transition between the subgiants and red giants. It is also unclear whether these formulations would agree with recent inferences of the internal rotation of subgiants in \citet{BuldgenMCMC2024}. There is also to this day no clear justification as to why the instability would switch from the behaviour described by \citet{spr02} to that of \citet{Fuller2019} at the end of the main-sequence.

Regarding the transport of chemical elements, we are either left with a too efficient mixing and large depletion of lithium and beryllium in the standard formulation of the magnetic Tayler instability or no depletion at all in the case of the more efficient formalisms used when studying red giants and subgiants. The remaining question is whether an in-between regime could be derived and whether some free parameters can be tweaked in the equations to ensure a higher depletion of lithium in the new formalisms and a steep decrease of the transport coefficients allowing to simultaneously reproduce the observed beryllium depletion. 

Turning to the transport equations used in stellar evolutionary models, the only free parameter to tweak is the efficiency of the horizontal turbulence, noted $\rm{D_{h}}$, intervening in both the transport by the shear instability and the meridional circulation. The transport of chemicals by both these processes is modelled as diffusive, using the following expressions in \citet{Eggenberger2022}.

The diffusive transport of chemicals by rotation is given by the following expression:
\begin{align}
\rm{D_{\rm{Rota}}}=\rm{D_{eff}}+ \rm{D_{X}},\label{eq:Deff1}
\end{align}
with $\rm{D_{eff}}$ the transport by the meridional circulation and $\rm{D_{X}}$ the transport by shear.
For the meridional circulation, $\rm{D_{eff}}$ is defined as
\begin{align}
\rm{D_{eff}}=\frac{\vert r U(r) \vert^{2}}{30\rm{D_{h}}}, \label{eq:Deff2}
\end{align}
with $\rm{D_{h}}$ the horizontal turbulence coefficient and $U(r)$ the velocity of the meridional circulation. Both $\rm{D_{h}}$ and $U(r)$ are related through some analytical prescriptions from \citet{Zahn1992}, \citet{Maeder2003} and \citet{Mathis2004} that we recall below. 
\begin{align}
\rm{D_{h}}&=\frac{1}{c_{h}} r \vert 2 V(r)-\alpha U(r) \vert, \label{eq:Meridional1}\\
\rm{D_{h}}&=A r (r \bar{\Omega}(r) V(r) \vert 2 V(r) - \alpha U(r) \vert )^{1/3}, \label{eq:Meridional2}\\
\rm{D_{h}}&= \left(\frac{\beta}{10} \right)^{1/2} (r^{2} \bar{\Omega}(r))^{1/2} \left( r  \vert 2 V(r) - \alpha U(r) \vert \right)^{1/2}, \label{eq:Meridional3}
\end{align}
with $U(r)$ and $V(r)$ the vertical and horizontal components of the velocity of the meridional circulation, $\bar{\Omega}(r)$ the average angular velocity over latitude at a given radius, $\alpha=\frac{1}{2}\frac{d\ln r^{2} \bar{\Omega}}{d \ln r}$, while $c_{h}$, $A$ and $\beta$ are constants. 

Regarding the shear instability, we follow the prescription of \citet{Talon1997} 
\begin{align}
\rm{D_{X}}\approx \frac{2 Ri_{c}(dW/dz)^{2}}{N^{2}_{T}/(K+\rm{D_{h}})+N^{2}_{\mu}/\rm{D_{h}}}, \label{eq:ShearTalon}
\end{align}
with $\rm{D_{h}}$ the horizontal turbulence coefficient, $Ri_{c}$ the critical Richardson number, $dW/dz=r \sin\theta (\rm{d}\Omega/\rm{d}r)$, the shear rate, K the thermal diffusivity and $N_{\mu}$ and $N_{T}$ the chemical and thermal contribution to the Brunt-V\"ais\"al\"a frequency. 

From a direct comparison between the asymptotic coefficients, we note that the transport coefficient of chemical elements in the revised magnetic instability formalism is about $10^4$ times smaller than the one from the original formulation. This explains why such models behave essentially as standard solar models from a chemical point of view. The key points is that, as noted by \citet{Maeder2004}, the magnetic Tayler instability itself does not transport lots of chemicals and indeed, the observed depletion of lithium and beryllium is due to the effect of shear and meridional circulation, as regulated by the magnetic instability to the critical value of shear above which the magnetic Tayler instability operates. It is thus natural that a more efficient magnetic instability leads to a lower degree of shear and thus a lower overall mixing. The remaining question is whether an arbitrary increase in $\rm{D_{h}}$ may compensate for this, keeping in mind that intrinsic treatments of the transport of chemicals by rotation might slightly alter the observed trends \citep{Dumont2021}. 

First, we can comment on the required increase. An increase of 4 orders of magnitude already cast doubts on the reliability of the magnetic Tayler instability to be considered responsible for the lithium and beryllium depletion in the Sun. Indeed, the horizontal turbulence as derived from the above expressions may lead to values of one or two orders of magnitude at most \citep[see e.g.][for an application to massive stars]{Nandal2024}. Nevertheless, one could be under the impression that increasing $\rm{D_{h}}$ would be the solution, due to the fact that a direct increase of $\rm{D_{h}}$ increases $\rm{D_{X}}$ in Eq. \ref{eq:ShearTalon} and thus the mixing by the shear instability which could remain localized. However, the increase of $\rm{D_{h}}$ would also increase the efficiency of the transport by the meridional circulation. Indeed, in each of the expressions, we see that 
\begin{align}
U(r) \propto \frac{1}{\alpha} \rm{D_{h}}^{n}
\end{align}
with $n \geq 1$. This is also in line with \citet{Chaboyer1992} and Eq. 4.37 and 4.38 in \citet{Maeder1998}. In other words, $\rm{D_{eff}}$ would increase under both the effect of the increase in $\rm{D_{h}}$ and the reduced shear induced by the efficient transport by the magnetic instability, implying that $\alpha \ll 1$. Therefore, such an increased transport by the meridional circulation would imply mixing chemicals down to temperatures high enough to lead to the efficient combustion of beryllium.

In essence, the meridional circulation naturally becomes more efficient if the horizontal turbulence is increased, meaning that simultaneous reproduction of both lithium and beryllium may not be achieved by modifying $\rm{D_{h}}$. In other words, the magnetic Tayler instability, in its current implementation or calibration to reproduce observational data, cannot be linked with the light elements depletion in the Sun. This is due to its intrinsic inefficient transport of chemicals, leaving either other magnetic instabilities \citep[e.g.][when talking about magnetic instabilities]{Jouve2020,Griffiths2022,Meduri2024}, improved depictions of the magnetic Tayler instability \citep[e.g.][]{Petitdemange2023, Petitdemange2024} taking into account additional hydrodynamical effects or other physical mecanisms \citep[][]{Charbonnel2005,Pincon2016}, as remaining candidates. It is however worth noting that the light element depletion in the \citet{Charbonnel2005} approach is due to the shear-layer oscillation and not the transport by the gravity waves themselves. In practice, this requires further investigations as to the actual impact of the solar tachocline and its potential link with the observed lithium and beryllium depletion.

Regarding the magnetic Tayler instability, its eligibility as an angular momentum transport candidate in the Sun is not affected by the conclusions of our analysis, but would require a measurement of the rotation of the deep solar core using gravity modes \citep{Appourchaux2010,Leibacher2023} to be confirmed beyond any doubts, as well as further theoretical developments beyond calibrations of its efficiency to explain the observed trends in asteroseismic inferences. It is also worth mentioning that there is no guarantee that the current picture drawn from asteroseismic inferences requires a unique solution. For example, the presence of a convective core during extended periods of evolution may also influence the presence of magnetic fields inside the star, thus the angular momentum transport in the interior and consequently the observed light element depletion. In the context of actual measurements of large radial fields \citep{Li2022,Li2023,Hatt2024}, incompatible with the magnetic Tayler instability, it might be worth considering the possibility of multiple explanations or other angular momentum transport candidates. 

\section{Conclusion}\label{Sec:Conc}

In this paper, we have investigated in detail the required properties of macroscopic mixing at the base of the solar convective envelope to reproduce simultaneously the observed lithium \citep{Wang2021} and beryllium \citep{Amarsi2024} photospheric abundances. We used solar calibrated models including both a simple parametric coefficient for macroscopic transport from \citet{Proffitt1991} and an asymptotic formulation based on models including angular momentum transport \citep{Eggenberger2022}. We also investigate the impact of replacing the base of the convective envelope at the helioseismic value on the inferred efficiency of macroscopic transport. We find that not taking into account the current position of the BCZ may lead to an overstimation of the transport by a factor of two and that the observed lithium and beryllium abundances imply an efficient transport with a relatively short extent, in line with what is found in more massive stars \citep{Richer2000}. This behaviour is impossible to reproduce with the current implementations of the effects of shear, meridional circulation and magnetic Tayler instability which intrinsically lead to an efficient mixing down to higher temperatures. We investigated the possibility to use revised prescriptions of these effects based on the current issues with angular momentum transport after the main-sequence and found that such models presented almost no mixing, as a result of the reduced shear allowed by the revised magnetic Tayler instabilities \citep{Fuller2019, Eggenberger2022TS}. Our findings thus cast doubts on the ability to link the observed light element depletion to the combined effects of shear, circulation and magnetic Tayler instability as suggested in \citet{Eggenberger2022}.

Our final calibration, using a power law of density as in \citet{Proffitt1991} leads to values of $n$ in Eq. \ref{eq:Proff} comprised between 3 and 6, the efficiency of the transport at the BCZ varying depending on this exact value but being of about $6000\; \rm{cm^{2}/s}$ for moderate values of $n$ that are in line with the values found for AMFM stars \citep{Richer2000}. 

Further analyses on solar-like stars observed by \textit{Kepler} might provide additional insights as to the underlying physical mechanism behind the light element depletion. In this respect the \textit{Kepler} Legacy sample \citep{Lund2017,Silva2017} might provide the perfect testbed as helium abundance in the envelope of some stars may be inferred from asteroseismic data and correlated with lithium and beryllium depletion (when these are available). Indeed, \citet{Verma2017} have already shown the need to introduce additional turbulent mixing  \citep{Verma2019}. The availability of surface or even internal rotation for some of these stars make them prime targets to push further our understanding of the missing dynamical processes in our current generation of stellar models. Due to their importance as tracers of mixing and the key role of lithium in primordial nucleosynthesis models, further extension of our analysis to low-metallicity regimes and globular clusters would also provide additional tests of the prescriptions we use for the Sun in a more global context of the theory of stellar structure and evolution. In this respect the role of the PLATO mission \citep{Rauer2024} in providing additional high-quality asteroseismic targets for which high-quality spectroscopic follow-up will be crucial. Such stars would allow us to place basic requirements for the underlying processes leading to the observed light element depletion and their potential link to the efficient angular momentum transport processes acting in the solar and stellar interiors.

\section*{Acknowledgements}

We thank the referee for their careful reading of the manuscript. GB acknowledges fundings from the Fonds National de la Recherche Scientifique (FNRS) as a postdoctoral researcher. AMA gratefully acknowledges support from the Swedish Research Council (VR 2020-03940) and from the Crafoord Foundation via the Royal Swedish Academy of Sciences (CR 2024-0015). We acknowledge support by the ISSI team ``Probing the core of the Sun and the stars'' (ID 423) led by Thierry Appourchaux. 

\bibliography{biblioarticleBe}

\end{document}